\documentclass[twocolumn,epjc3]{svjour3}
\smartqed  
\RequirePackage{graphicx}

\journalname{Eur. Phys. J. C}
\usepackage{color}
\usepackage{amssymb}

\begin{document}

\title{Quasiperiodic oscillations around hairy black holes in Horndeski gravity}

\author{Javlon~Rayimbaev \thanksref{e1, addr11,addr12,addr13,addr14} \and Konstantinos F. Dialektopoulos \thanksref{e2, addr21, addr22} \and Furkat~Sarikulov \thanksref{e4,addr30,addr31,addr33} \and Ahmadjon~Abdujabbarov \thanksref{e3,addr13,addr14,addr31} }
\thankstext{e1}{javlonrayimbaev6@gmail.com}
\thankstext{e2}{kdialet@gmai.com}
\thankstext{e3}{ahmadjon@astrin.uz}
\thankstext{e4}{furqatsariquloff@gmail.com}

\institute{Institute of Fundamental and Applied Research, National Research University TIIAME, Kori Niyoziy 39, Tashkent 100000, Uzbekistan\label{addr11} \and
Akfa University,  Milliy Bog Street 264, Tashkent 111221, Uzbekistan \label{addr12} \and 
National University of Uzbekistan, Tashkent 100174, Uzbekistan \label{addr13} \and 
Tashkent State Technical University, Tashkent 100095, Uzbekistan \label{addr14}\and
%
Department of Physics, Nazarbayev University, 53 Kabanbay Batyr Avenue, 010000 Astana, Kazakhstan \label{addr21} \and 
Laboratory of Physics, Faculty of Engineering, Aristotle University of Thessaloniki, 54124 Thessaloniki, Greece \label{addr22} \and 
School of Mathematics and Natural Sciences, New Uzbekistan University, Mustaqillik Ave. 54, Tashkent 100007, Uzbekistan\label{addr30} \and 
Ulugh Beg Astronomical Institute, Astronomy str. 33, Tashkent 100052, Uzbekistan \label{addr31} \and
Inha University in Tashkent, Ziyolilar 9, Tashkent 100170, Uzbekistan \label{addr33} 
}

\date{Received: date / Accepted: date}

\maketitle

\begin{abstract}

Testing gravity theories and their parameters using observations is an important issue in relativistic astrophysics. In this context, we investigate the motion of test particles and their harmonic oscillations in the spacetime of non-rotating hairy black holes (BHs) in Hordeski gravity, together with astrophysical applications of quasiperiodic oscillations (QPOs). We show possible values of upper and lower frequencies of twin-peak QPOs which may occur in the orbits from innermost stable circular orbits to infinity for various values of the Horndeski parameter $q$ in relativistic precession, warped disk models, and three different sub-models of the epicyclic resonant model. We also study the behaviour of the QPO orbits and their position relative to innermost stable circular orbits (ISCOs) with respect to different values of the parameter $q$. {It is obtained that at a critical value of the Horndeski parameter ISCO radius takes $6M$ which has been in the pure Schwarzschild case.} Finally, we obtain mass constraints of the central BH of microquasars GRS 1915+105 and XTE 1550-564 at the GR limit and the possible value of the Horndeski parameter in the frame of the above-mentioned QPO models. The analysis of orbits of twin peak QPOs with the ratio of upper and lower frequencies 3:2, around the BHs in the frame of relativistic precession (RP) and epicyclic resonance (ER4) QPO models have shown that the orbits locate close to the ISCO. The distance between QPO orbits and ISCO is obtained to be less than the error of the observations.

\end{abstract}



\section{Introduction}
\label{introduction}

Quasiperiodic oscillations (QPOs) are astrophysical phenomena corresponding to (several) peaks observed in radio-to-X-ray bands of the electromagnetic spectrum. Twin-peaked QPOs in microquasars can be observed through the process of matter accreting into neutron stars, white dwarfs, or black holes (BHs)~\cite{Ingram2016MNRAS,Stuchlik2013AA,Stella1998ApJL,Rezzolla_qpo_03a}. One may distinguish two types of such twin peak QPOs: high frequency (HF) corresponding to the frequency range from 0.1 to 1 kHz and low frequency (LF) with a frequency less than 0.1 kHz. 

The QPOs with several peaks can be observed in low-mass X-ray binaries (LMXBs) where one (or both) component(s) consists of neutron stars~\cite{Ingram2010MNRAS,Schaab1999MNRAS,Torok2005AA}. Spectral analysis has also shown that twin-peaked QPOs and QPOs with several peaks differ from each other. Therefore, separate models must be considered to describe the origin of QPOs in different scenarios ~\cite{Rezzolla_qpo_03b,Germana2017PhRvD}. Spectral behaviour of QPOs and their temporal variability may be helpful in determining/measuring magnetic field properties in the accretion disk of neutron stars and BHs in microquasars~\cite{Torok2019MNRAS,Zdunik2000AA,Klis2000ARAA}. 

Despite the fact that there are many research works devoted to the high-accuracy measurements of the QPO frequencies and testing gravity theories, the unique astrophysical model explicitly describing the behaviour of QPOs has not been yet proposed. 
One of the promising models for explaining the generation of QPOs is connected with the dynamics of test particles around BHs and their harmonic oscillations in the radial, vertical, and azimuthal directions. Thus, using data from QPOs detected in microquasars~\cite{Bambi17e,Stuchlik2015MNRAS} one may test the spacetime around BHs.  
Moreover, investigations of QPO models based on the orbital motion of test particles might enable us to study the inner edge of the accretion disk surrounding BHs. Our previous studies have shown that QPO orbits are close to the innermost stable circular orbit (ISCO) of particles \cite{Stuchlik2011AA,Torok2011AA,Rayimbaev2021Galax}. 
This implies that using the QPO analysis, one may estimate the values of the mass of the central BH and its spin (or/and other parameters). Moreover, the studies are also helpful in determining which theory of gravity plays the dominant role in spacetime around the central BHs of the microquasars \cite{Silbergleit2001ApJ,Wagoner2001ApJL,Rayimbaev2022PDU,Vrba2021Univ,Vrba2021EPJP,Vrba2021JCAP,Rayimbaev2021EPJCQPO,Rayimbaev2021GalaxQPO,Rayimbaev2021PhRvDQPO,Franchini2017PhRvD,Maselli2017ApJ}.

Being well tested and justified in weak and strong field regimes, general relativity needs to be further updated/modified in order to resolve its shortcomings, such as the nature of the dark sector, singularities, the value of the cosmological constant, the $H_0$ tension and more. A very bright example is that general relativity meets the singularity issue during gravitational collapse, and it requires the presence of so-called dark energy to explain the accelerated expansion of the present universe~\cite{Clifton:2011jh,2015Emanuela}. 


Among the various ways to modify general relativity, scalar tensor theories such as the Horndeski theory of gravity \cite{Horndeski:1974wa} reflect a special interest in astrophysics due to their advantages. Particularly, one of the interesting features of the theory is that even though its Lagrangian contains second-order derivatives of the metric and the scalar field, the equations of motion contain up to second-order derivatives of the metric and a scalar field (see, e.g.~~\cite{2019Kobayashi}).
The field equations of the Horndeski theory have been obtained using the variation principle applied to the action containing the metric and a scalar field (which has a form of scalar-tensor models having Galilean symmetry in flat spacetime~\cite{Nicolis:2008in}) and contain all symmetries of general relativity ~\cite{Damour:1992we,Horbatsch:2015bua,Capozziello:2018gms}.  The effects of Horndeski gravity in strong field regimes near gravitating compact objects~\cite{Maselli:2016gxk} and in large cosmological scales~\cite{Kase:2018aps} have been extensively studied. The gravitational lensing by BHs in Horndeski's theory in weak field limits has been investigated by the authors of Ref. \cite{Ali2021IJGMM}. Furthermore, Horndeski gravity has been formulated in the teleparallel geometry \cite{Bahamonde:2019shr}, and apart from the richer phenomenology that the theory presents, it has been shown that terms that were severely constrained from GW170817, can be revived in this framework \cite{Bahamonde:2019ipm,Bahamonde:2021dqn}. Other applications in the teleparallel framework can be found in Refs.~\cite{Bahamonde:2020cfv,Dialektopoulos:2021ryi} as well as in the reviews \cite{Bahamonde:2021gfp,CANTATA:2021ktz}.

{BHs in Horndeski gravity have a nontrivial scalar field profile, which is commonly called hair.} BH solutions within Horndeski gravity have been obtained in Refs. \cite{Rinaldi:2012vy,Babichev:2014fka,Babichev:2017guv,Anabalon:2013oea,Cisterna:2014nua,Bravo-Gaete:2014haa}, particularly, the solutions with  a radially dependent hairy scalar field have been studied in Refs.~\cite{Sotiriou:2013qea,Sotiriou:2014pfa,Babichev:2016rlq,Benkel:2016rlz}. Further analysis of the theory and corresponding solutions of Horndeski gravity have been intensively studied in Refs.~\cite{khoury,Bergliaffa:2021diw}.

Additionally, testing Horndeski gravity using observational data from the size of the shadow of rotating supermassive BH M87* by EHT collaboration, has been widely studied in \cite{AfrinGhosh2022ApJ} and relationships between the spin of the BH and the Horndeski parameters have been obtained. Authors in Ref.~\cite{Kumar2022EPJC} have provided detailed analyses on gravitational lensing by Horndeski black holes and applied the calculations to several astrophysical supermassive BHs. Moreover, studies of mass to radius relation of neutron stars within Horndeski gravity have been investigated in Ref. \cite{Maselli2016PhRvD}. 

In this paper, we plan to study the motion of the test particles around the hairy BH and its application to describe the QPOs. The paper is organized as follows: In Sect.~\ref{Sec:metric1} we review the hairy BH solution. Sec.~\ref{Sect3} is devoted to studying the motion of the test particles around the hairy BH in Horndeski gravity. The fundamental frequencies associated with circular orbits of the particles have been studied in Sect.~\ref{Sec4}. The application of particle motion and fundamental frequencies to QPO analysis has been provided in Sect.~\ref{Sec5}. We conclude our results in Sec.~\ref{Sec:conclusion}. 

Throughout this paper, we use the (--, +, +, +) signature for the spacetime metric and system of units where $G=1=c$.   

\section{Hairy black hole in Horndeski gravity}\label{Sec:metric1}

Horndeski gravity is a modification of general relativity, being the most general scalar-tensor theory in four dimensions that leads to second-order field equations \cite{Horndeski:1974wa,2019Kobayashi}. Its action is described by
{\begin{equation}\label{action}
    \mathcal{S} = \int d^4x \sqrt{-g} \sum _{i=2}^5 L_i\,,
\end{equation}}
where $L_i$ are the following Lagrangian and \\$X = -\frac{1}{2}\nabla _\mu \phi \nabla ^\mu \phi$
\begin{eqnarray}
L_2 &=& G_2(\phi,X)\ , \quad L_3 = -G_3(\phi,X){\square}\phi\, , \\ \nonumber
L_4 &=& G_4(\phi,X){R} + G_{4,X}(\phi,X)\Big[({\square}\phi)^2 
\\ &-& {\nabla}{}_{\mu}{\nabla}_{\nu}\phi{\nabla}^{\mu}\nabla^{\nu}\phi \Big]\, ,\\ \nonumber
L_5 &=& G_5(\phi,X){G}_{\mu\nu}{\nabla}^{\mu}{\nabla}^{\nu}\phi - \frac{1}{6}G_{5,X}(\phi,X) \\ \nonumber &\times & \Big[({\square}\phi )^3 + 2{\nabla}_{\nu}{\nabla}_{\mu}\phi{\nabla}^{\nu}{\nabla}^{\lambda}\phi{\nabla}_{\lambda}{\nabla}^{\mu}\phi \\ &-& 3{\square}\phi {\nabla}{}_{\mu}{\nabla}_{\nu}\phi{\nabla}^{\mu}{\nabla}^{\nu}\phi\Big]\ . \label{LC_Horn_Comp4}
\end{eqnarray}
$G_i (\phi, X)$ are arbitrary functions of the scalar field and its kinetic term. Following \cite{Bergliaffa:2021diw} we study a subclass of the above theory that considers $G_i$ to be only functions of the kinetic term, i.e. $G_i(X)$. In addition, $G_5(X) = 0$.
%

{The field equations can be obtained by varying the action (\ref{action}) with respect to the metric,
\begin{equation}\label{eq:metricEq}
    G_4(X)G_{\mu\nu}=T_{\mu\nu}
\end{equation}
where
\begin{eqnarray}
\nonumber
    T_{\mu\nu} &=& \frac{1}{2}\left(G_{2,X}\nabla _{\mu} \phi \nabla _{\nu} \phi + G_2 g_{\mu\nu}\right)\\ \nonumber & +& \frac{1}{2} G_{3,X}\Big( \nabla _{\mu} \phi \nabla _{\nu} \phi \square \phi - g_{\mu\nu} \nabla _\alpha X \nabla ^\alpha \phi \\ \nonumber &+& 2\nabla _{(\mu} X \nabla _{\nu)} \phi \Big) - G_{4,X} \Big[ \nabla _\gamma \nabla _\mu \phi \nabla ^\gamma \nabla _\nu \phi-  \\ \nonumber
     &-& \nabla _\mu \nabla _\nu \phi  \square \phi + \frac{1}{2} g_{\mu\nu} \Big( (\square \phi)^2 - (\nabla _\alpha \nabla _\beta \phi )^2 \\ \nonumber &-& 2 R_{\sigma \gamma} \nabla ^{\sigma} \phi \nabla ^{\gamma} \phi \Big) -\frac{R}{2} \nabla _\mu \phi \nabla _\nu \phi \\ \nonumber & +& 2R_{\sigma (\mu|} \nabla ^{\sigma} \phi \nabla _{|\nu)} \phi  + R_{\sigma \nu\gamma \mu} \nabla ^\sigma \phi \nabla ^\gamma \phi\Big] - G_{4,XX}  \\ \nonumber &\times& \Big[ g_{\mu\nu} \left( \nabla _\alpha X \nabla ^\alpha \phi \square \phi +\nabla _\alpha X \nabla ^\alpha X\right) + \frac{1}{2} \nabla _\mu \phi \nabla _\nu \phi \\ \nonumber &\times& \left((\nabla _\alpha  \nabla _\beta \phi)^2 - (\square \phi)^2 \right) - \nabla _\mu X \nabla _\nu X \\ \nonumber &-& 2\square \phi \nabla _{(\mu} X\nabla_{\nu )}\phi - \nabla _\gamma X \\ &\times& \left( \nabla ^\gamma \phi \nabla _\mu \nabla _\nu \phi - 2 \nabla ^\gamma \nabla _{(\mu}\phi \nabla _{\nu)}\phi\right) \Big].
\end{eqnarray}
}

{The finite four-current $j^\mu$, which identifies the invariance of the scalar field under shift symmetry can be defined as, 
\begin{equation}
    j^\mu=\frac{1}{\sqrt{-g}}\frac{\delta {\cal S}}{\delta \phi _{,\mu}}\,.
\end{equation}
and in our case, it reads
\begin{eqnarray}
\nonumber
    j^{\nu} = &-& G_{2,X} \phi ^{,\nu} - G_{3,X} (\phi ^{,\nu}\square \phi + X^{,\nu}) \\ \nonumber &-& G_{4,X} (\phi^{,\nu}R - 2R^{\nu\sigma}\phi_{,\sigma}) \\ \nonumber &-& G_{4,XX} \Big[\phi^{,\nu}\left( (\square\phi)^2 - \nabla_\alpha \nabla_\beta \phi \nabla^\alpha \nabla^\beta \phi\right) \nonumber \\
    &+&2 \left(X^{,\nu} \square X - X_{,\mu}\nabla^{\mu}\nabla^{\nu}\phi \right) \Big]\ .
\end{eqnarray}
For the metric
\begin{equation}
    ds^2 = - A(r) dt^2 + \frac{1}{B(r)}dr^2 + r^2 d\Omega ^2\ ,
\end{equation}
the non-vanishing component of the above current takes the form
\begin{eqnarray}\label{eq:j^r}
   j^r &=& -G_{2,X} B \phi ' - G_{3,X} \frac{4A+rA'}{2rA}B^2 \phi^{'2} \\ \nonumber &+& 2 G_{4,X} \frac{B}{r^2 A}\left[  (B-1)A + r BA'\right]\phi' \\ \nonumber & - & 2 G_{4,XX} \frac{B^3 (A+rA')}{r^2A}\phi^{'3}\,,
\end{eqnarray}
where $'$ denotes differentiation with respect to the radial coordinate.}

{For simplicity, we set 
\begin{eqnarray}
    G_2 &=& \alpha _{21} X + \alpha _{22}(-X)^{\omega _2}\,, \\
    G_3 &=& \alpha _{31}(-X)^{\omega _3}\,,\\
    G_4 &=& \frac{1}{8\pi} + \alpha _{42}(-X)^{\omega _4}\,. 
\end{eqnarray}
For hairy solutions to exist, we set
\begin{equation}
    \alpha _{21} = \alpha _{31}  = 0\,,\quad \omega _2 = \frac{3}{2}\,, \quad \omega _4 = \frac{1}{2}\,,
\end{equation}
and for imposing $j^r = 0$ we obtain
\begin{equation}\label{phi_sol}
    \phi' = \pm \frac{2}{r}\sqrt{\frac{-\alpha _{42}}{3 B \alpha _{22}}}\,.
\end{equation}
In order to see the complete derivation, check \cite{Bergliaffa:2021diw}. From the metric equations \ref{eq:metricEq} we get
\begin{equation}\label{bh_metric}
    A(r) = B(r) = 1-\frac{2 M}{r}+\frac{q}{r}\ln \frac{r}{2 M},
\end{equation}
with $q$ being a constant
\begin{equation}
    q = \left(\frac{2}{3}\right)^{3/2} \kappa ^2 \alpha _{42}\sqrt{-\frac{\alpha_{42}}{\alpha_{22}}}\,.
\end{equation}
For the scalar field to satisfy the energy conditions, the expression in the square root in $q$ (and thus Eq.~\ref{phi_sol}) should be positive definite, otherwise, the scalar field would be imaginary. This means that $\alpha _{42}$ has to be negative or $\alpha _{22}$, but not both at the same time. 
}

\begin{figure}[ht!]
\centering
\includegraphics[width=0.45\textwidth]{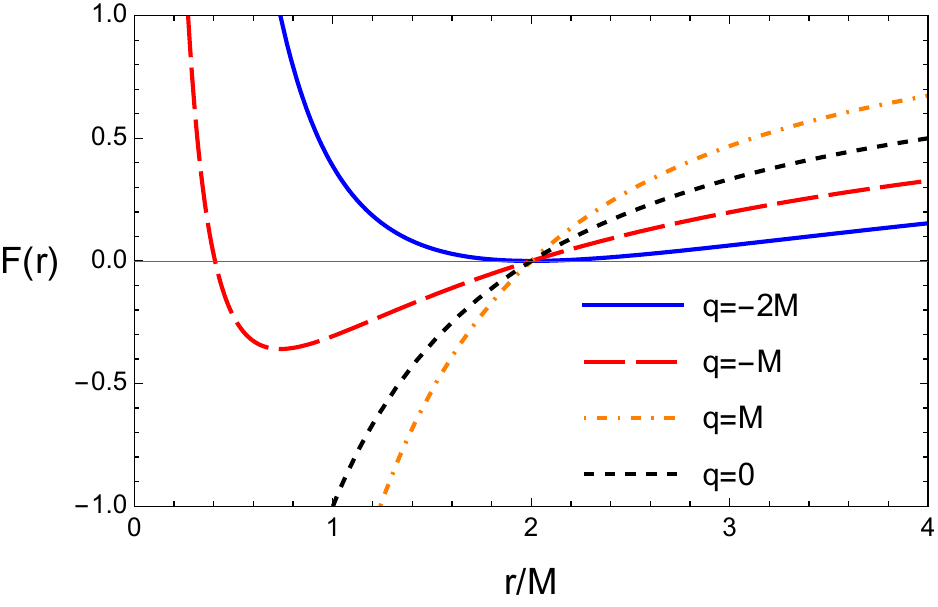}
  \caption{The radial dependence of the lapse function for the different values of $q$ in Horndeski gravity. The zeros of the lapse function determined the horizon of the BH.
  \label{grr}}
\end{figure}

Summarizing, the geometry around a hairy BH in Horndeski gravity can be described by the following spacetime
\begin{eqnarray}\label{metric}
ds^2=-F(r)dt^2+\frac{1}{F(r)}dr^2+r^2(d\theta^2+\sin^2\theta d\phi^2),
\end{eqnarray}
with the metric function defined as 
\begin{equation}\label{Ffunc}
F(r)=1-\frac{2 M}{r}+\frac{q}{r}\ln \frac{r}{2 M},
\end{equation}
where $M$ is the BH mass and {$q$ is the scalar charge with the dimension of length related to the non-trivial scalar field}. From Fig.~\ref{grr} it is clearly seen that the metric (\ref{metric}) always has the horizon at the Schwarzschild radius ($r=2M$) irrespective of the value of the parameter $q$. Moreover, BH has two horizons when $-2<q/M<0$. 

\begin{figure}
\centering
\includegraphics[width=0.45\textwidth]{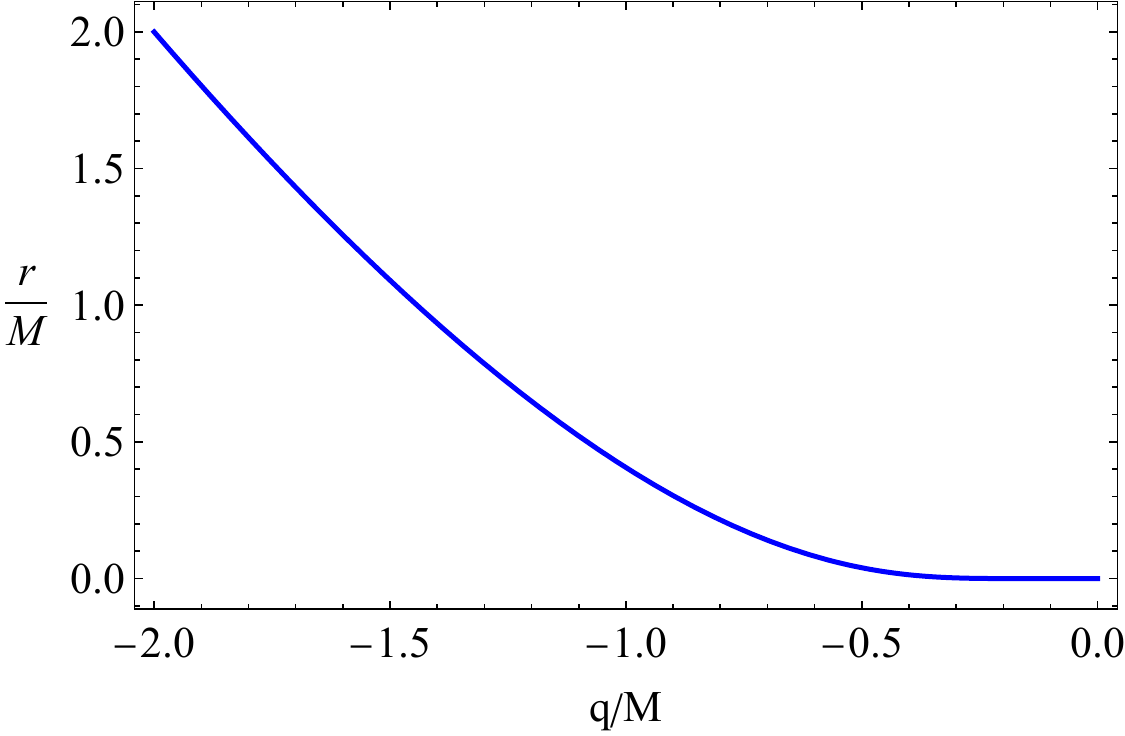}
 \caption{The Cauchy horizon of a hairy BH in Horndeski gravity as a function of the parameter $q$.\label{2hor}}
\end{figure}

The event horizon radius of the BH can be found using $g_{rr} \to \infty$ or, say, $g^{rr}=0$ which reduces $F(r)=0$. One can easily see from the expression of $F(r)$ that there are two event horizons in the spacetime of the hairy BH: outer and inner. The outer one is the event horizon and the inner one is called the Cauchy horizon. The outer horizon is 2$M$ for the values of the parameter $q/M$ from -2 to 0, and the inner horizon vanishes at $q=0$ and at $q=-2M$ and these two horizons coincide with each other at $r=2M$. With an increase in the parameter $q$, the Cauchy horizon decreases (see Fig.~\ref{2hor}). The expression of the inner horizon has the form,
\begin{equation}
r=q\ {\rm ProductLog} \left(\frac{2M}{q} e^{\frac{2M}{q}}\right),
\end{equation}
where  ProductLog$(z)$, for arbitrary $z$, is defined as the principal solution of the equation $We^W = z$.

\section{Test particle motion around hairy black hole \label{Sect3}}

In this section, we consider the dynamics of electrically neutral test particles around a hairy BH in Horndeski gravity using the following Lagrangian for the test particles
\begin{eqnarray}
    {\cal L}_p = \frac{1}{2} m g_{\mu\nu} \dot{x}^\mu \dot{x}^\nu.
\end{eqnarray}
where $m$ is the mass of the test particle {and overdot stands for the derivative with respect to proper time $\tau$. It is worth noting that  $x(\tau)$ is the particle worldline, parametrized by the proper time $\tau$ and the particle's four-velocity, $u_\mu$ is defined as $u^\mu = dx^\mu/d\tau$. }
Due to the symmetry of the spacetime around a spherically symmetric BH, one may directly obtain two integrals of motion: energy ${\cal E}$ and angular momentum $\cal L$ in the form
\begin{eqnarray}
\label{Ekilling}
{\cal E}&=&-u_{\mu} \xi^{\mu}, \qquad \dot{t}=\frac{\cal E}{F(r)}\ ,\\
\label{Lkilling}
{\cal L}&=&u_{\mu} \eta^{\mu}, \qquad \dot{\phi}=\frac{\cal L}{r^2 \sin^2{\theta}}\ ,
\end{eqnarray}
where $\xi^{\mu}$ and $\eta^{\mu}$ are the Killing vectors associated with time-translation and 
rotational invariance, respectively. ${\cal E}=E/m$ and ${\cal L}=L/m$ in Eqs.~(\ref{Ekilling})-(\ref{Lkilling}) stand for specific energy and angular momentum. Equations of motion for the test particle are then governed by the normalization condition
\begin{equation}
\label{ncon}
g_{\mu\nu}u^\mu u^\nu=\varepsilon\ ,
\end{equation}
where $\varepsilon$ equals 0 and -1 for massless and massive particles, respectively.

For the massive particles' equation of motion governed by timelike geodesics of spacetime and the equations of motion can be found by using Eq.~\ref{ncon}.
Taking into account Eqs.~(\ref{Ekilling})-(\ref{Lkilling}) one may obtain the equations of motion in the separated and integrated form as
\begin{eqnarray}
    \dot{r}^2 = {\cal E} - F(r) \Big(1+\frac{{\cal K}}{r^2}\Big), \ \ 
    \dot{\theta}^2 = \frac{1}{g_{\theta\theta}^2}\Big({\cal K}-\frac{{\cal L}^2}{\sin^2{\theta}}\Big)\ ,
\end{eqnarray}
where ${\cal K}$ denotes the Carter constant corresponding to the 
total angular momentum.

Restricting the motion of the particle to a constant plane, in which $\theta ={\rm const}$ and $\dot{\theta}=0$, thus, the Carter constant takes the form ${\cal K} = {\cal L}^2/\sin^2{\theta}$ and the equation of the radial motion can be expressed in the form:
\begin{eqnarray}
\dot{r}^2={\cal E}^2-V_{\rm eff}\ ,
\end{eqnarray}
where the effective potential of the radial motion reads
\begin{eqnarray}
\label{Veff}
V_{\rm eff} = F(r) \left(1+\frac{{{\cal L}^2}}{r^2 \sin ^2\theta }\right). 
\end{eqnarray}

Now, we apply standard conditions for the circular motion, which corresponds to zero radial velocity $\dot{r} = 0$ and acceleration \"{\it r} = 0. One can obtain the expressions of the specific angular momentum and the specific energy for circular orbits at the equatorial plane  ($\theta = \pi/2$) in the following form:
\begin{eqnarray}
{\cal L}^2 = r^2 {\frac{2 M-q \left(\ln \frac{r}{2 M}-1\right)}{q \left(3 \ln \frac{r}{2 M}-1\right)+2 r-6 M}}\ , \\
{\cal E}^2 = \frac{2\left(q \ln \frac{r}{2 M}+r-2 M\right)^2}{{r \left[q \left(3 \ln \frac{r}{2 M}-1\right)+2 r-6 M\right]}}\ .
\end{eqnarray}

\begin{figure}[ht!]
\centering
\includegraphics[width=0.45\textwidth]{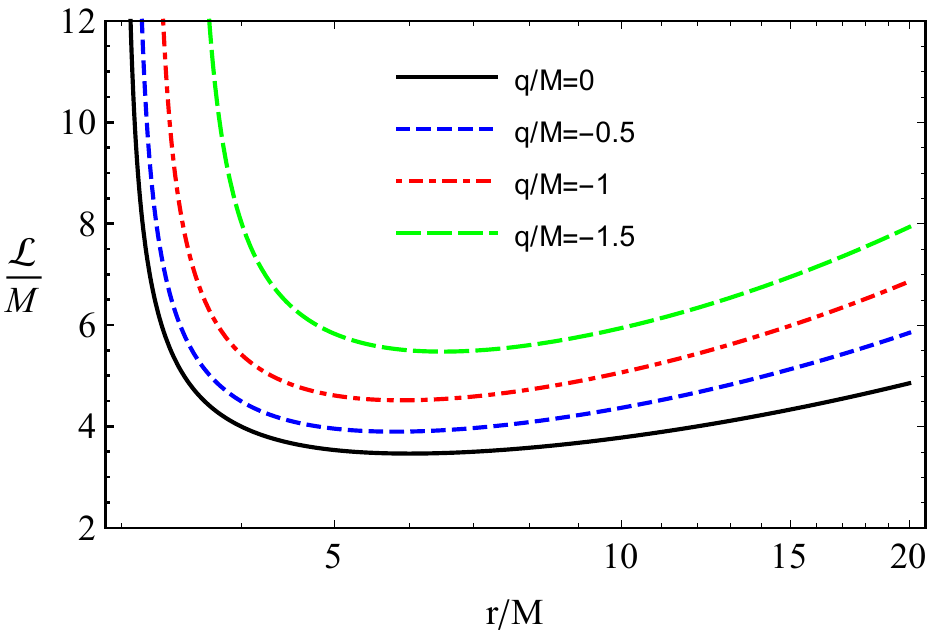}
 \includegraphics[width=0.45\textwidth]{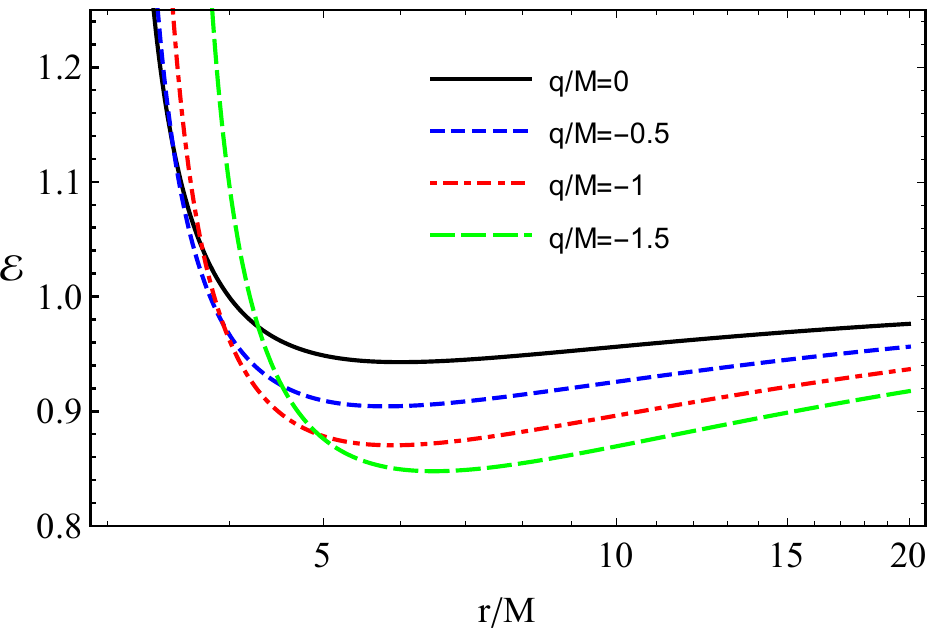}
 \caption{The radial dependence of specific angular momentum (top panel) and energy (bottom panel) of the test particle in circular orbits for the different values of parameter $q$. \label{LandE}}
\end{figure}

Figure \ref{LandE} demonstrates radial profiles of specific energy and angular momentum
of test particles along circular stable orbits around a hairy BH in Horndeski gravity, for the different values of parameter $q$. It is seen from the figure that the specific angular momentum and  its minimum value increase due to the presence of the parameter $q$. On the other hand, specific energy decreases with the decrease of parameter $q$. Note that the metric (\ref{metric}) reduces to Schwarzschild when $q = 0$.

One can easily see from Fig.~\ref{LandE} that the value of the radius of marginally stable circular orbits of test particles, which may be also referred to as the radius of the photonsphere, increases with the increase of the parameter $q$. In order to determine the radius of the photonsphere we solve the radial geodesic equation,
\begin{equation}
    q \left(1-3 \ln \frac{r}{2 M}\right)+6 M-2 r=0\,.
\end{equation}

\begin{figure}[h!]
\centering
\includegraphics[width=0.45\textwidth]{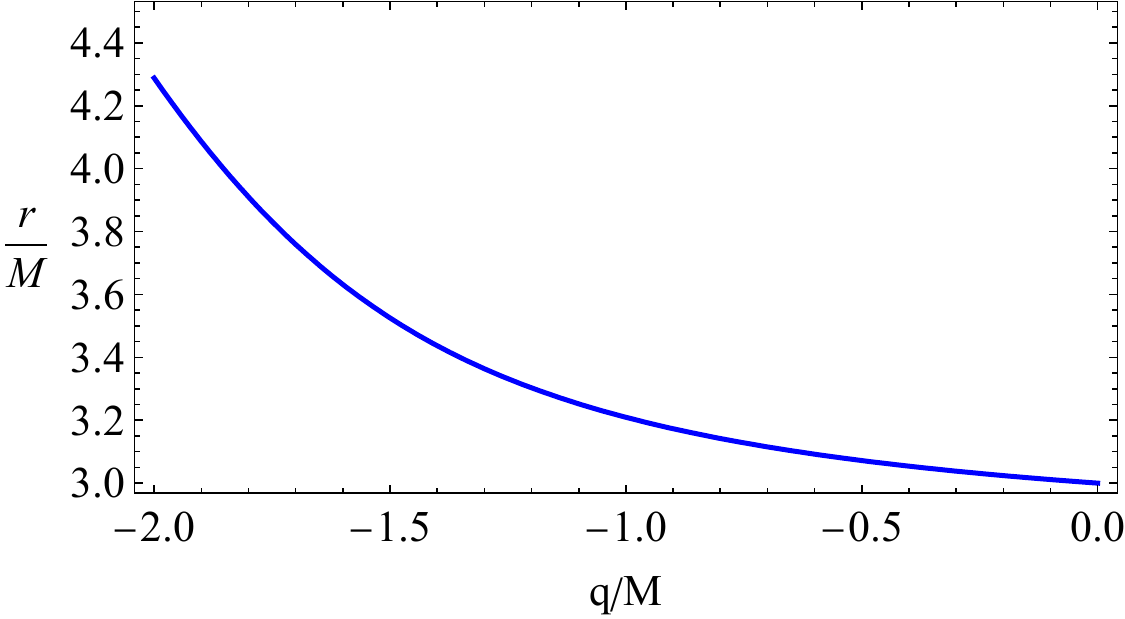}
 \caption{The dependence of the radius of photonsphere around hairy BHs in Horndeski gravity from the parameter $q$. \label{phdep}}
\end{figure}

In Fig.~\ref{phdep} we provide the photonsphere radius around hairy BHs in Horndeski gravity as a function of the parameter $q$. {One may easily see from the figure that as the parameter $q$ goes from $-2M$ to zero, the radius of the photonsphere decreases from about $4.3M$ to $3M$.} 

\subsection{Innermost stable circular orbits}

The stable circular orbits occur at the radius $r=r_{min}$ where the minimum of the effective potential takes place. The innermost stable circular orbit corresponds to \\$\partial_{rr}V_{\rm eff}=0$ and/or $\partial_{r}{\cal{L}}=0$ which leads to the same results. After some algebraic simplifications, the equation for the ISCO radius of test particles is obtained in the following form:
\begin{eqnarray}\label{iscoeq} \nonumber
&& 6 M+3 q+r-3 q \ln \frac{r}{2 M}\\ && +\frac{(q+r) (3 q+2 r)}{-3 q \ln \frac{r}{M}+6 M+q+q \ln (8)-2 r}=0\ .
\end{eqnarray}

\begin{figure}[h!]
\centering
\includegraphics[width=0.45\textwidth]{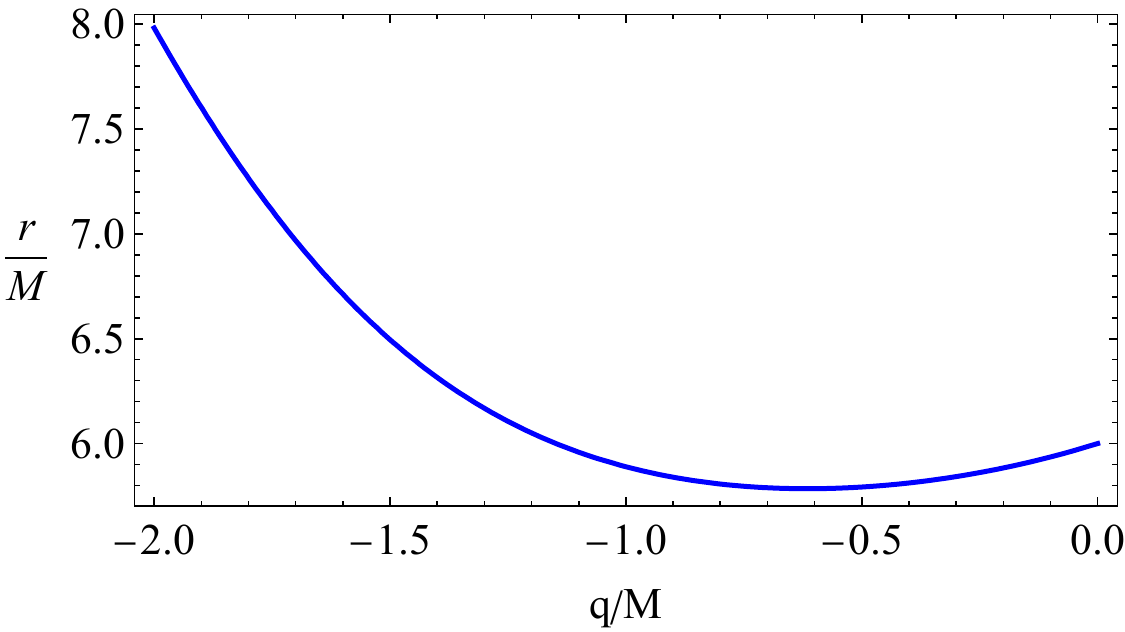}
 \caption{The dependence of ISCO radius of test particles around hairy BHs in Horndeski gravity from the $q$ parameter. \label{iscodep}}
\end{figure}

As we mentioned above that the solution of this equation with respect to radial coordinates implies ISCO radius. However, due to the complicated form of Eq.~(\ref{iscoeq}) it is hard to solve it analytically with respect to $r$. In order to analyse the behaviour of ISCO radius, we present the numerical results of the dependence of ISCO radius $r_{\rm ISCO}$ from $q/M$ in Fig.~\ref{iscodep}.

Figure~\ref{iscodep} represents the dependence of ISCO radius from the parameter $q$.  It is observed from Fig.~\ref{iscodep} that the radius of ISCO, first, decreases with increasing parameter $q$, reaches the minimum value, ($r_{\rm ISCO})_{\rm min} \approx 5.7846 M$ at $q/M=-0.615$ and then increases again up to $6M$. For $q=0$ and $q=-1.14M$, the ISCO radius equals to $6M$, {which covers the result for the Schwarzschild BH case.} This implies that in these values of the parameter $q$ it has a degeneracy behaviour. 

\section{Fundamental frequencies \label{Sec4}}

In this section, we provide derivations of an expression for the fundamental frequencies governed by the particle orbiting around the hairy BH in Horndeski gravity. In particular, we explore frequencies of Keplerian orbits and the radial \& vertical (to the orbital plane) oscillations, which are helpful in investigations of QPO models.  


\subsection{Keplerian frequencies}
The angular velocity of the particles orbiting around the BH measured by an observer located at infinity is called the Keplerian frequency $\Omega_K=d\phi/dt$ and can be expressed as
\begin{equation}
    \Omega_K = \sqrt{-\frac{\partial_r g_{tt}}{\partial_r g_{\phi\phi}}}=\sqrt{\frac{F'(r)}{2r}}.
\end{equation}

The expression of the frequency in the Horndeski spacetime metric given in Eq.(\ref{metric}) takes the following form
\begin{eqnarray}
    \label{Kepleq}
    \Omega_K^2= \frac{M}{r^3}+\frac{q}{2r^3}\left(1-\ln \frac{r}{2M}\right) .
\end{eqnarray}

{Furthermore, to express the frequencies in Hz we use the following equation:
\begin{equation}\label{nuHzs}
    \nu_{K,r,\theta}=\frac{c^3}{2\pi GM}\Omega_{K,r,\theta}\ ,
\end{equation}
where $c$ and $G$ are the speed of light in a vacuum and the gravitational (Newtonian) constant.}

\begin{figure}
\centering
\includegraphics[width=0.45\textwidth]{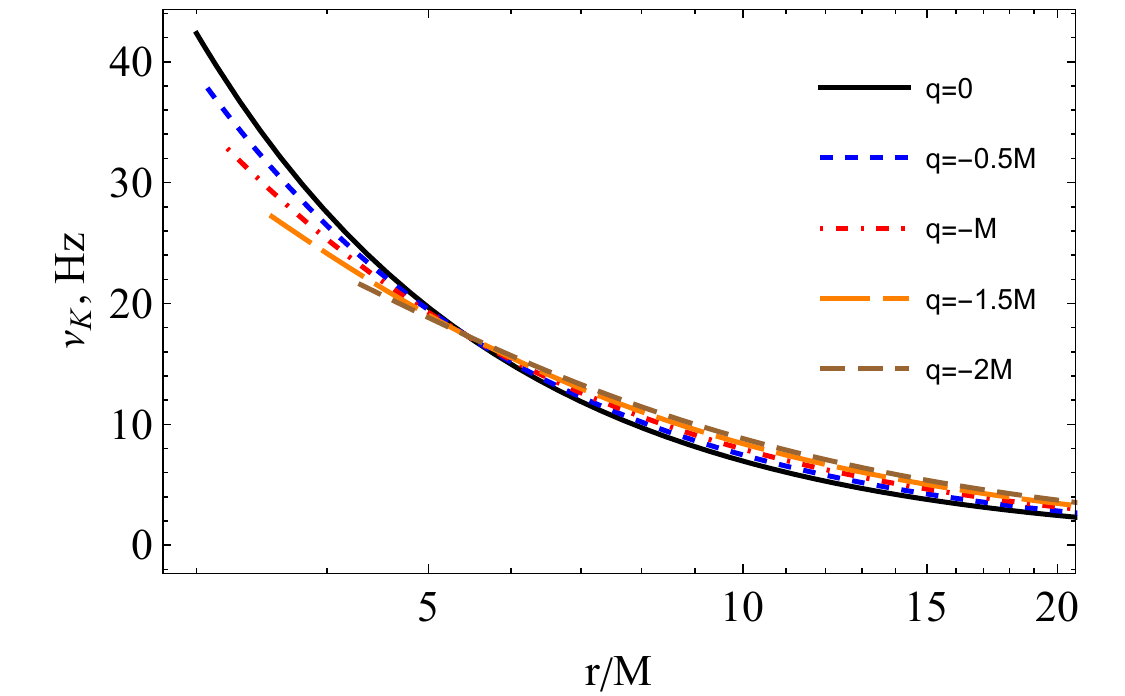}
 \caption{Radial dependence of frequencies of particles in Keplerian orbits around hairy BHs in Horndeski gravity for the different values of the parameter $q$.  \label{kepl}}
\end{figure}

Figure~\ref{kepl} demonstrates radial profiles of the frequencies of Keplerian orbits of test particles around hairy BHs in Horndeski gravity for different values of the parameter $q$. It is observed that the increase of $q$ causes to decrease of the Keplerian frequency up to the distance of about $(4.43-4.45)M$. However, far from this range, the frequency decreases slower due to the presence of the parameter $q$.

\subsection{Harmonic oscillations}

We consider a test particle to oscillate along the radial, angular, and vertical axes in its stable orbits around a static BH in the equatorial plane due to the small displacement from the orbits as $r_0+\delta r$ and $\pi/2+\delta \theta$. One can calculate the frequencies of the radial and vertical oscillations measured by a distant observer using harmonic oscillator equations~\cite{Bardeen68}:  
\begin{eqnarray}
\frac{d^2\delta r}{dt^2}+\Omega_r^2 \delta r=0\ , \qquad \frac{d^2\delta\theta}{dt^2}+\Omega_\theta^2 \delta\theta=0\ ,   
\end{eqnarray}
where 
\begin{eqnarray}
\Omega_r^2=-\frac{1}{2g_{rr}(u^t)^2}\partial_r^2V_{\rm eff}(r,\theta)\Big |_{\theta=\pi/2}\ ,
\\ \Omega_\theta^2=-\frac{1}{2g_{\theta\theta}(u^t)^2}\partial_\theta^2V_{\rm eff}(r,\theta)\Big |_{\theta=\pi/2}\ ,
\end{eqnarray}
are the frequencies of the radial and vertical oscillations, respectively. After some algebraic calculation and simplifications, we immediately have expressions for the frequencies in the spacetime of static BHs expressed as follows~\cite{Rayimbaev2021Galax}:

\begin{eqnarray}
\label{omegar}\nonumber 
 \Omega_r^2&=&\Omega^2_K\Bigg(1-\frac{6 M}{r} \\ &+&\frac{q}{r} \left[3 \ln\frac{r}{2 M}-\frac{r-2M-q \ln ({r}/{2 M})}{q \left(1+\ln ({r}/{2 M})\right)+2 M}\right]\Bigg) \ ,
\\ \label{omegatheta}
    \Omega_{\theta} &=& \Omega_{\phi} = \Omega_K\ .
\end{eqnarray}

\begin{figure}
\centering
\includegraphics[width=0.45\textwidth]{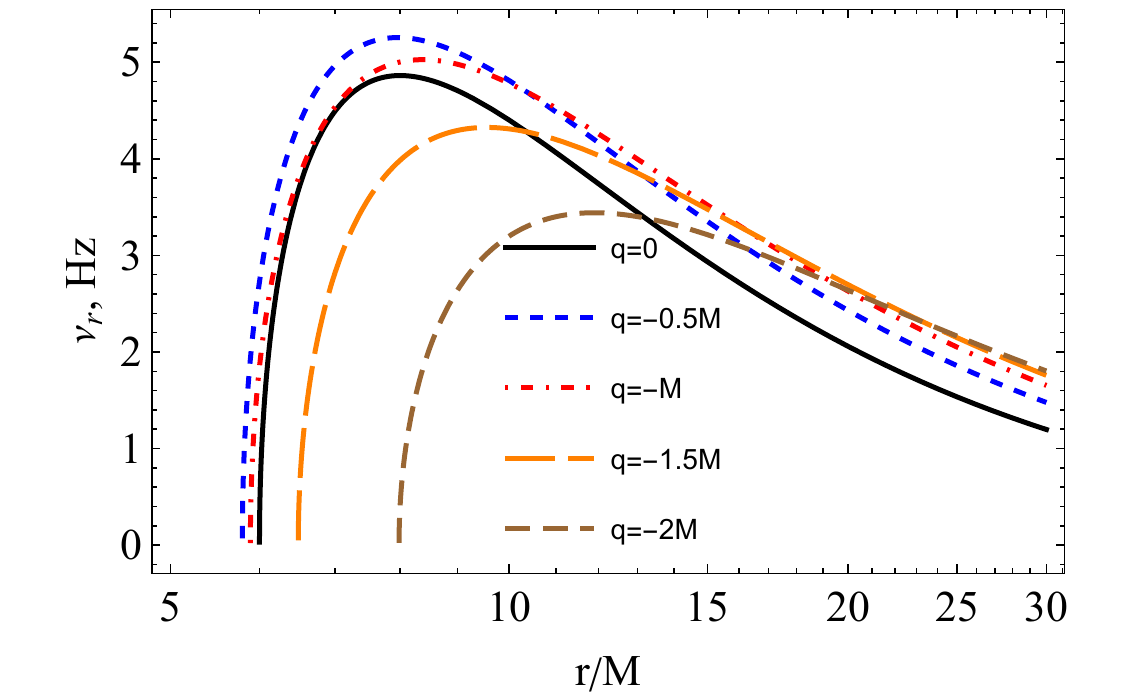}
 \caption{Radial dependence of frequencies of radial oscillations of test particles in stable circular orbits around the hairy BH in Horndeski gravity.  \label{radfreqfig}}
\end{figure}
The radial profiles of the frequencies of the radial oscillations of particles around a hairy BH in Horndeski gravity are shown in Fig.~\ref{radfreqfig} for the various values of the parameter $q$. It is found that the maximum value of the frequency increases with the decrease of $q$ up to $q=-0.6M$ and then decreases back.   

\section{Astrophysical applications: QPOs\label{Sec5}}

This section is devoted to exploring possible values of frequencies of twin-peak QPOs around hairy BH in Horndeski gravity using various QPO models, in particular, to compare $q$ parameter effects on the upper and lower frequencies with the effects of the spin of rotating Kerr BH~\cite{Stuchlik2016AA}. In addition, we also focus on determining the relationship between the mass of the hairy BH and the parameter $q$ using their observational frequency data from QPOs.  We also consider that the BHs at the center of the microquasars GRS 1915+105 \cite{Abramowicz2001AA} and XTE 1550-564 \cite{Remillard2002ApJ} are hairy ones.

\subsection{QPO models}

 In this subsection, we plan to study the possible values of the upper and lower frequencies by the following models for twin peak {HF} QPOs {described by the fundamental frequencies of test particles around compact gravitating objects ($\nu_{r,\theta,\phi}=c^3/(2\pi GM)\Omega_{r,\theta,\phi}$):} 

\begin{itemize}

\item Relativistic precession (RP) model has been proposed by Stella \& Vietri~\cite{Stella1998ApJL} for kHz twin peak QPOs corresponding to the frequencies in the range from 0.2 to 1.25 kHz from neutron stars in LMXRBs. Later, it has been shown that the model is also applicable to BH candidates in the binary systems of BH and neutron stars \cite{Stella2001AIPC}. RP model has been further developed by Ingram~\cite{Ingram2014MNRAS} in order to obtain precise measurements of the mass and spin of central BH in microquasars using data from the power-density spectrum of the BH accretion disk. According to the RP model, the upper and lower frequencies are described by the frequencies of the radial, vertical and orbital oscillations in the forms $\nu_U=\nu_\phi$ and $\nu_L=\nu_\phi-\nu_r$, respectively. 

\item The epicyclic resonance (ER) model considers resonances of the axisymmetric oscillation modes of a thin accretion disc around BHs~\cite{Abramowicz2001AA}. The frequencies of the disc oscillation modes are related to the frequencies of orbital and epicyclic oscillations of the circular geodesics of the test particles. Here, we use the variations of ER model: ER2, ER3, and ER4 which differ in their oscillation modes. The corresponding upper and lower frequencies in ER2-4 models, defined as $\nu_U=2\nu_\theta-\nu_r$ \& $\nu_L=\nu_r$, $\nu_U=\nu_\theta+\nu_r$ \& $\nu_L=\nu_\theta$  and  $\nu_U=\nu_\theta+\nu_r$ \& $\nu_L=\nu_\theta-\nu_r$, respectively ~\cite{Abramowicz2001AA}.

\item The warped disc (WD) model assumes non-axisymmetric oscillatory modes of a thin accretion disc around BHs and neutron stars~\cite{Kato2004PASJ,Kato2008PASJ}. In the WD model, the upper and lower frequencies are defined as $\nu_U=2\nu_\phi-\nu_r$,  $\nu_L=2(\nu_\phi-\nu_r)$ ~\cite{Kato2004PASJ,Kato2008PASJ}. The vertical oscillatory frequency $\nu_\theta$ has been introduced by the assumptions of vertical axial symmetric oscillations of the accretion disc that cause the disc to warp.
\end{itemize} 

\begin{figure*}[ht!]\centering
\includegraphics[width=0.32\textwidth]{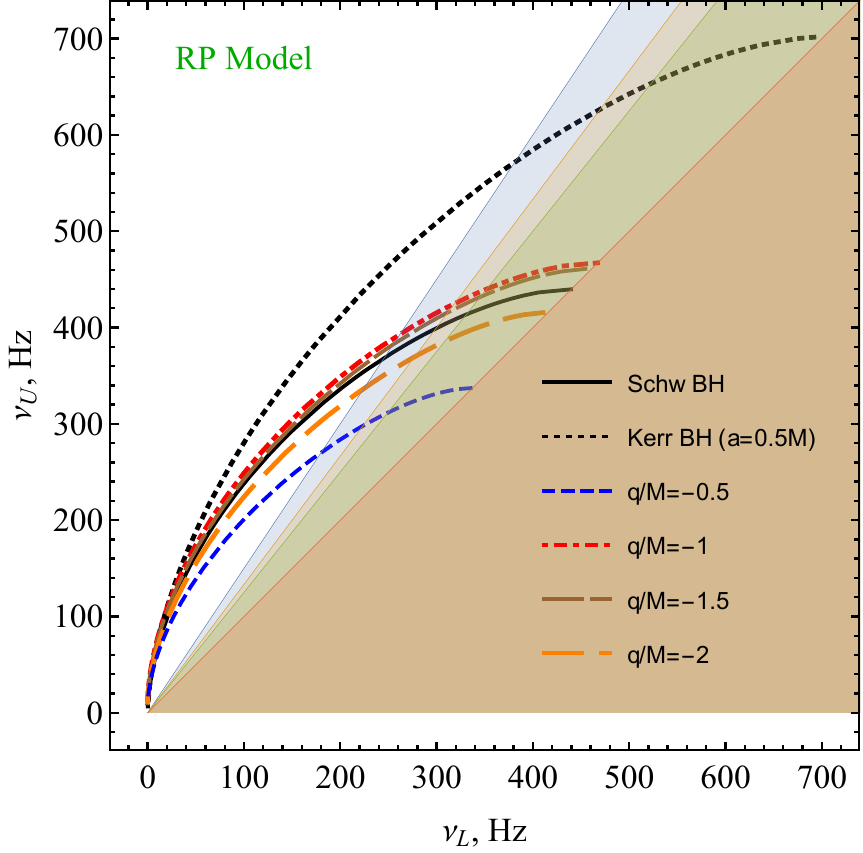}
 \includegraphics[width=0.32\textwidth]{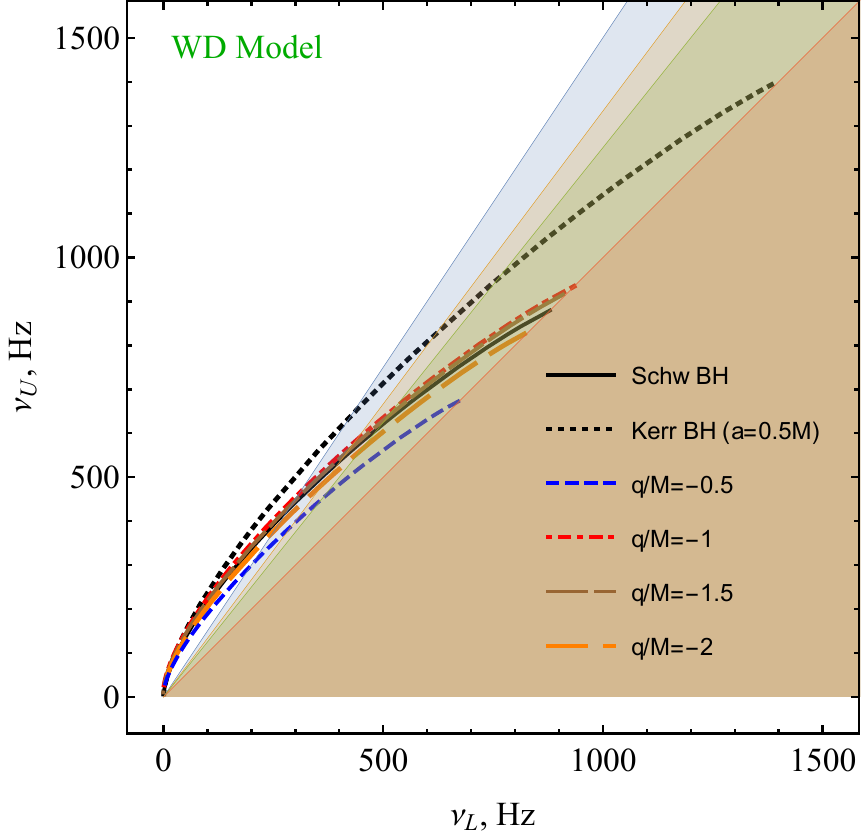}
 \includegraphics[width=0.32\textwidth]{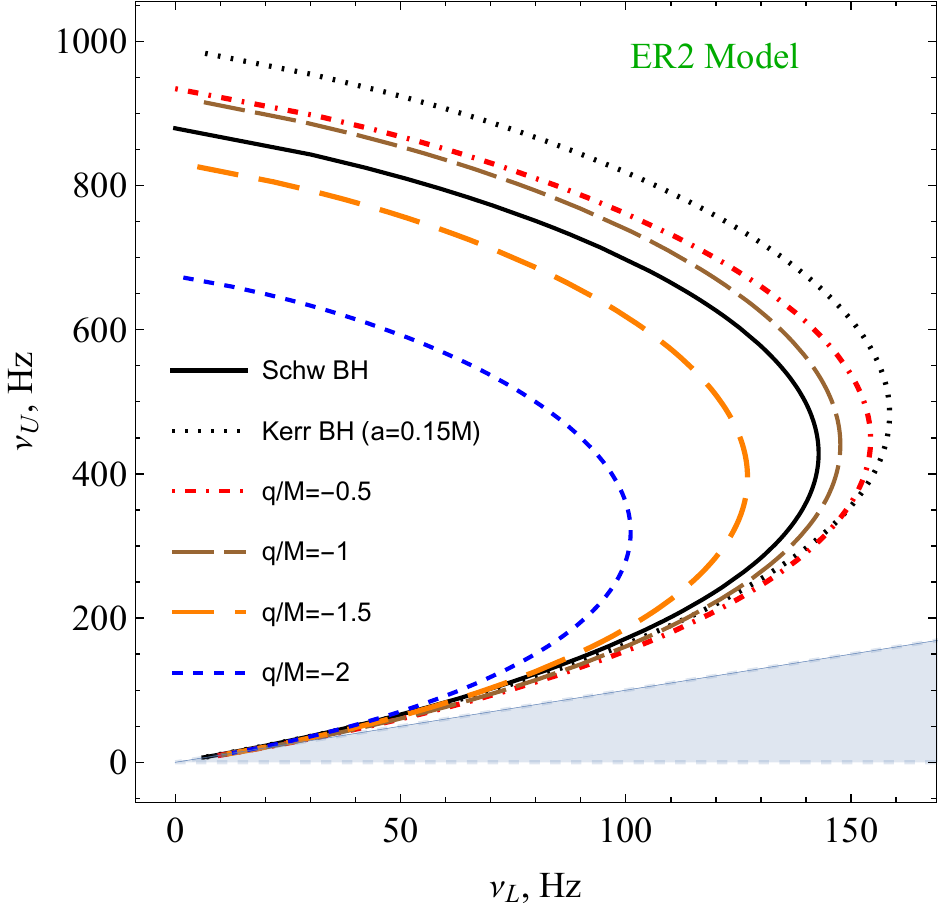}
 \includegraphics[width=0.32\textwidth]{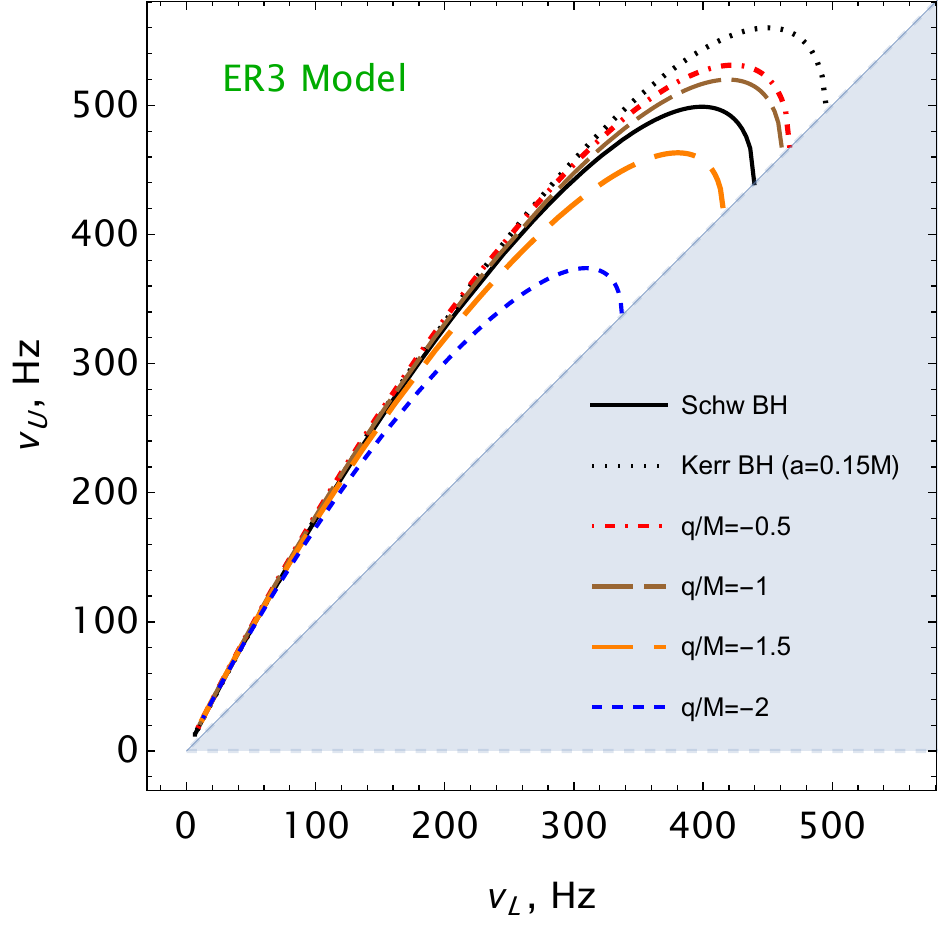}
 \includegraphics[width=0.32\textwidth]{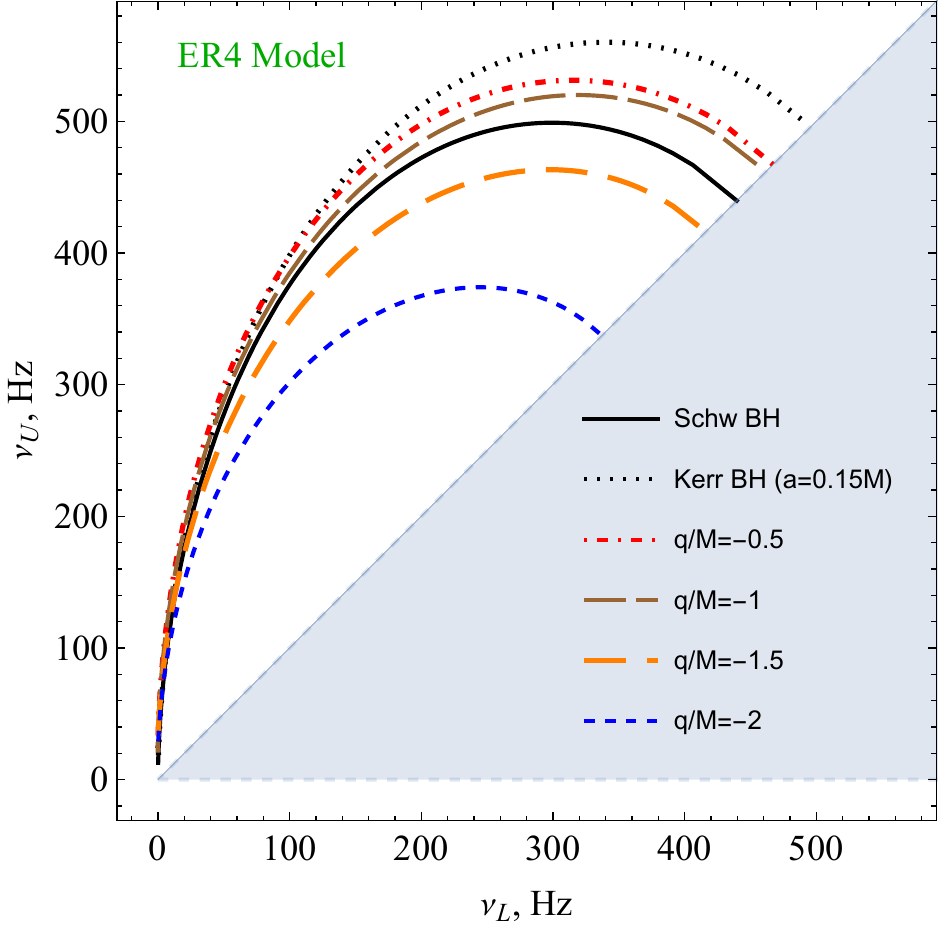}
 \caption{The relationship between the upper and lower frequencies of twin peak QPOs in the spacetime around hairy BHs in Horndeski gravity in the RP, ER and WD models for the different values of the parameter $q$. \label{Models}}
\end{figure*}

In Fig.~\ref{Models} we demonstrate the diagram $\nu_U-\nu_L$ for twin-peak QPOs around the hairy BHs in the RP, ER2-4, and WD models together with the comparisons of the QPOs around rotating Kerr BHs with the spin parameter $a/M=0.1$. In plotting the figure, we have taken the value of the central black hole mass $M=5M_\odot$ as a test stellar mass black hole. To obtain units of the frequencies in Hz we use Eq.(\ref{nuHzs}). The light-blue shaded area in the top-right and bottom panels and orange-shaded area in the top left and middle panels in the diagram imply the graveyard for twin peak QPOs. Any twin-peaked QPOs cannot be observed in that area. The inclined lines bordering the areas are deathlines for the twin-peak QPOs where the upper and lower frequencies are equal to each other and the two peaks in the twin-peak QPOs merge into a single peak. This implies that if a QPO position falls down under the deadline in the graveyard, then the QPO object disappears from observation. The diagram shows that the ratio of the upper and lower frequencies increases with negative values of the parameter $q$ is approximately $q\simeq-0.5M$. However, for values less than $q=-0.5M$ the ratio decreases.

Our numerical comparisons have shown that the parameter $q$ can mimic the spin of Kerr BH up to about $a=0.1M$ with its value $q\simeq-0.55M$ in all the models considered providing the same values for upper and lower frequencies in twin-peaked QPOs. This means that both a Kerr black hole and a hairy black hole can produce almost identical QPO frequencies by testing the surrounding particles. From this point of view, it is not possible to distinguish the two cases using the QPO analyzes. It requires, additionally, other independent types of observational data and theoretical detailed analysis. So, there is a degeneracy between the spin of Kerr black holes and the parameter of static hairy black holes. 

\subsection{QPO orbits}

In this subsection, we study relationships between the parameter $q$ and the radius of orbits where a QPO shines in the RP, ER2-4, and WD models by constructing the following equation for the ratio of upper and lower frequencies,
\begin{eqnarray}
\label{uplow32}
    3\nu_L(M,r,q)=2\nu_U(M,r,q)\ , \\ \label{uplow43} 4\nu_L(M,r,q)=3\nu_U(M,r,q)\ , \\ \label{uplow54} 5\nu_L(M,r,q)=4\nu_U(M,r,q)\ .
\end{eqnarray}

One may get the following equation for the relationships between the radius of orbits where the QPO appears with the ratio 3:2 and the parameter $q$, by substituting Eqs.~(\ref{nuHzs}), (\ref{omegar}) and (\ref{omegatheta}) into Eq.~(\ref{uplow32}) in RP model:
\begin{eqnarray}\label{rqQPO}
\frac{6 M}{r}-\frac{q}{r} \left[3 \ln\frac{r}{2 M}-\frac{r-2M-q \ln \frac{r}{2 M}}{q \left(1+\ln \frac{r}{2 M}\right)+2 M}\right]=\frac{2}{3}.
\end{eqnarray}
Due to the complex form of the Eq.~(\ref{rqQPO}) it is impossible to get the analytical expression for the radius. However, one can perform a numerical analysis of the dependence of the radius $r$ on $q$. Similar analyses on the QPO radius for the frequency ratios 4:3 \& 5:4 in WD \& ER QPO models can be performed. 

\begin{figure*}[ht!]\centering
\includegraphics[width=0.32\textwidth]{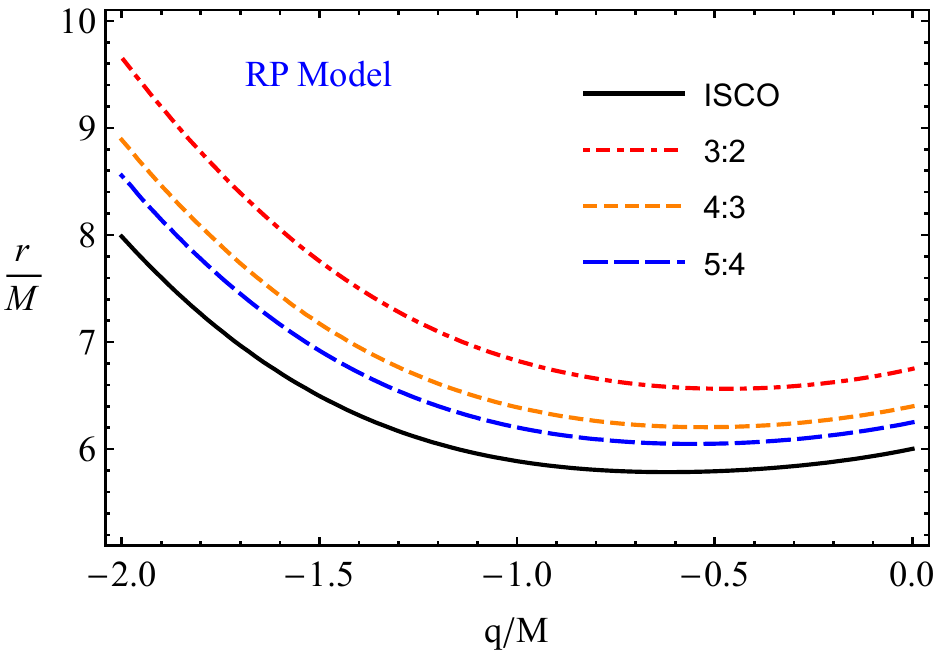}
\includegraphics[width=0.32\textwidth]{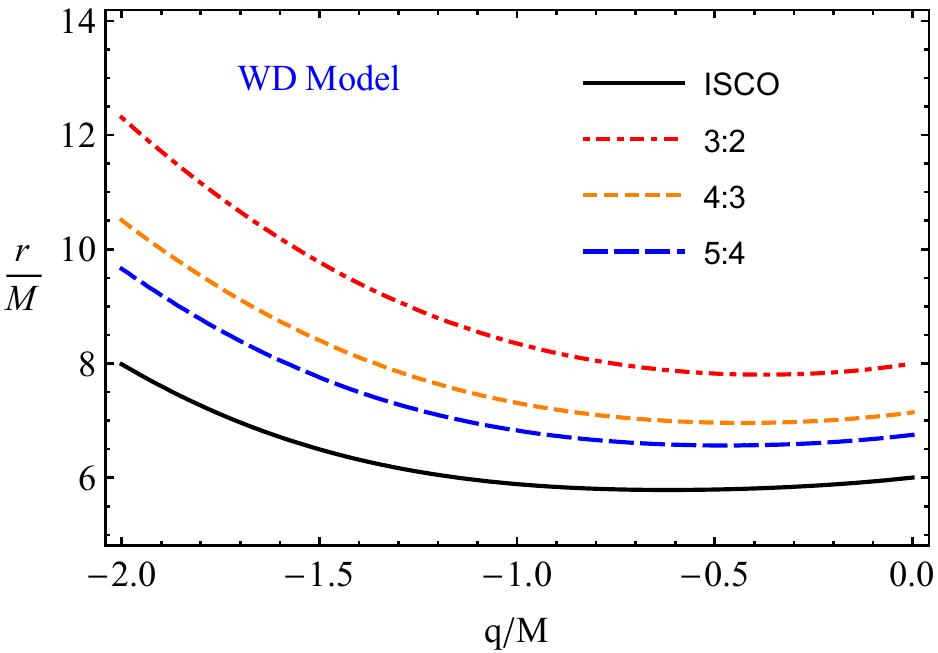}
\includegraphics[width=0.32\textwidth]{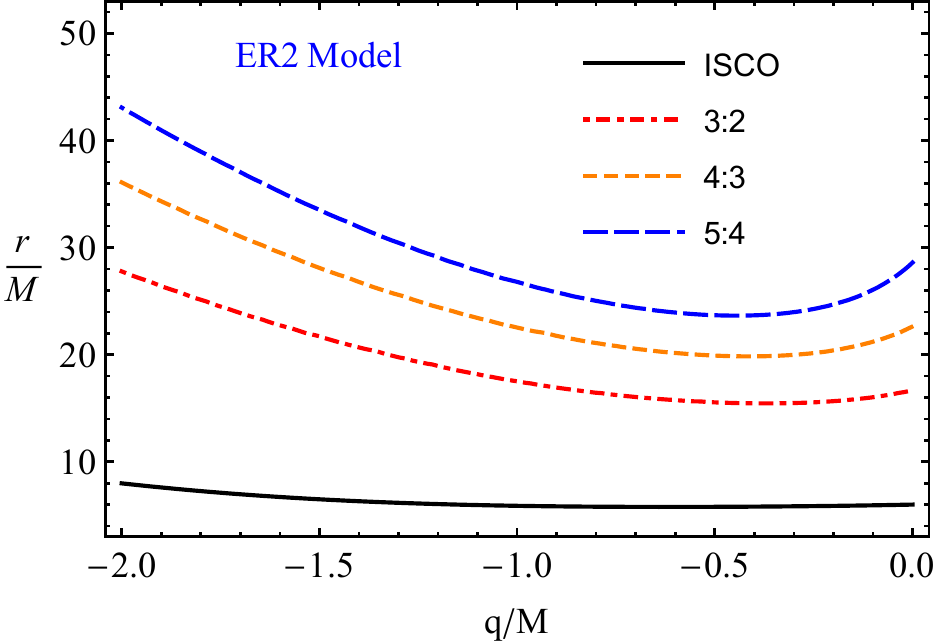}
\includegraphics[width=0.32\textwidth]{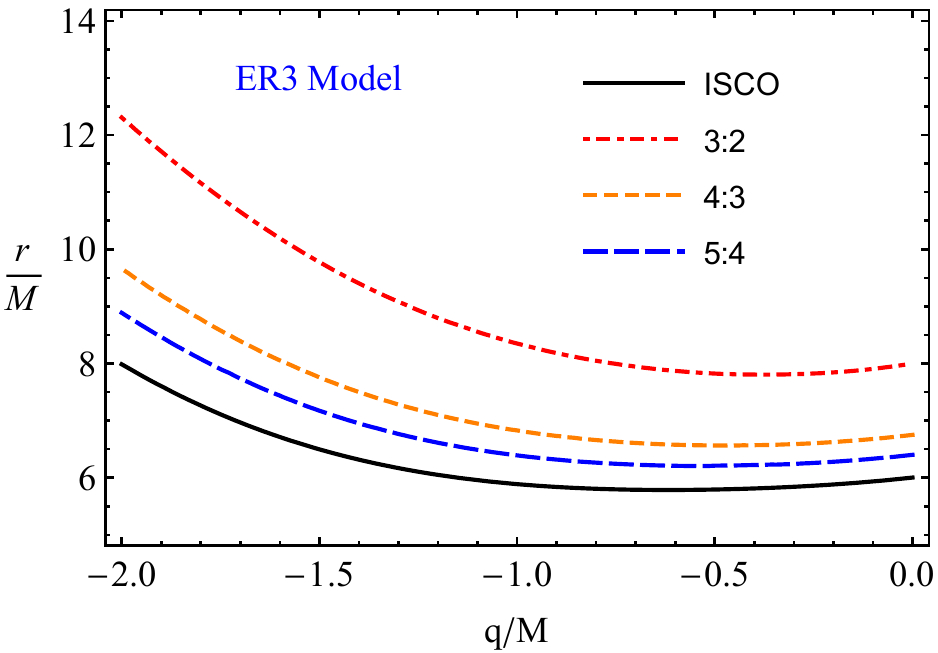}
\includegraphics[width=0.32\textwidth]{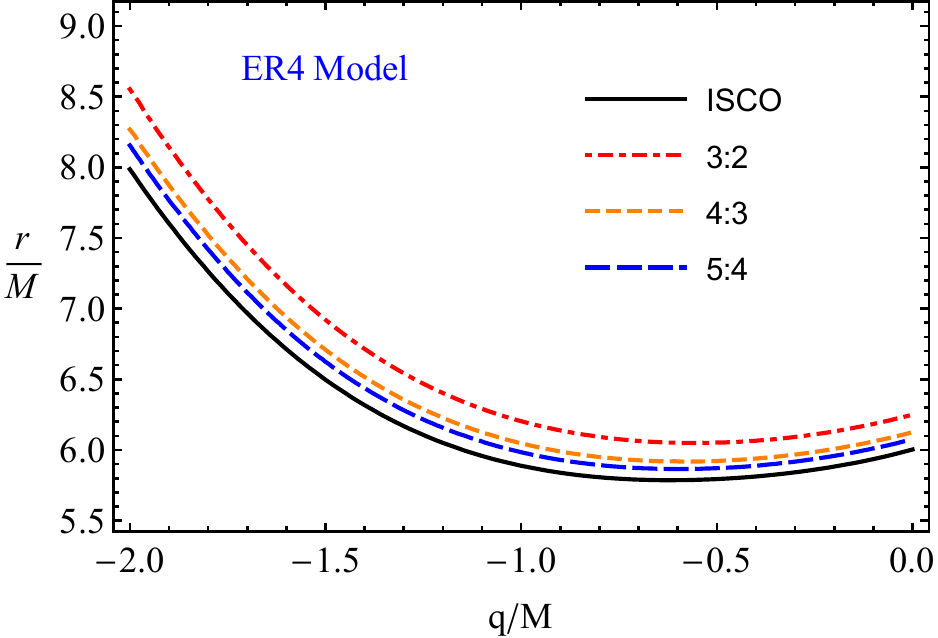}
\caption{The radius of orbits where twin peak QPOs are located with the ratios 3:2, 4:3, and 5:4 in RP, WD, and ER2-4 models and ISCO radius as a function of parameter $q$.
 \label{ISCOmodels}}
\end{figure*}

In Fig.~\ref{ISCOmodels} we demonstrate the QPO radius as a function of the parameter $q$. One can see from Fig.~\ref{ISCOmodels} that the QPO orbits are located outside the ISCO.
{One can see from Fig.~\ref{ISCOmodels} that the orbits of QPOs with 5:4 are closer to the central object than the other ratios. Thus, if a twin peak QPO shines in an orbit close to ISCO, the peak frequencies of the QPO become close to each other.} This means that if a twin peak QPO generates at ISCO by test particles, the two peaks unify and the upper and lower frequencies become equal to each other. 

Moreover, the distance between the QPO orbit and ISCO {depends on the QPO model and the Horndeski parameter $q$. In RP, WD, and ER3,4 models, the QPO orbits are located near ISCO. However, in the ER2 model, the orbits are quite far from the ISCO.} 

It is also observed from Fig.\ref{ISCOmodels} that the QPO profiles at $q/M=-0.5$ and -1 lie under the line corresponding to the Schwarzschild case ($q=0$), while in cases when $q/M=-1.5$ and $q/M=-2$ the lines appear under the Schwarzschild curve. One can explain the strange feature of QPO profiles using the similar behaviour of QPO orbits with ISCO \cite{Rayimbaev2022PDU,Rayimbaev2021GalaxQPO,Rayimbaev2022IJMPD1,Syunyaev1973}. One can see from Fig.\ref{iscodep} that there is a critical value in the Horndeski parameter, $q_{\rm cr}/M=-1.14834$ where ISCO radius equals $6M$. That implies this case, the Horndeski BH mimics the Schwarzchild one, providing the same ISCO radius as well as QPO orbits. When $|q|<|q_{\rm cr}|$ the ISCO around the Horndeski BH is less than $6M$ (see Fig.\ref{iscodep}) and QPO orbits come close to $6M$, positioning inside the QPO orbits in Schwarzschild case. It causes the ratio of QPO frequencies to be higher than the ratio in the Schwarzschild case. While at $q<q_{\rm cr}$ the QPO orbits lie further $6M$ where a twin peak QPO can be generated with frequencies smaller than the Schwarzschild case.

Now, we show the distance between the ISCO and QPO orbits ($\delta=r_{\rm QPO}-r_{\rm ISCO}$) for the selected the QPOs observed in the microquasars GRS 1915+105 and XTE 1550-564, assuming the central BH is a hairy BH in Horndeski gravity. The frequencies of the QPO sources in the microquasars GRS 1915+105 and XTE 1550-564 are $\nu_U=168\pm 5$ Hz \& $\nu_L=113\pm 3$ Hz, and $\nu_U=184 \pm 5$Hz \& $\nu_L=276 \pm 3$ Hz \cite{Vrba2021JCAP,Vrba2021EPJP,Abramowicz2001AA}.

\begin{figure}\centering
\includegraphics[width=0.420\textwidth]{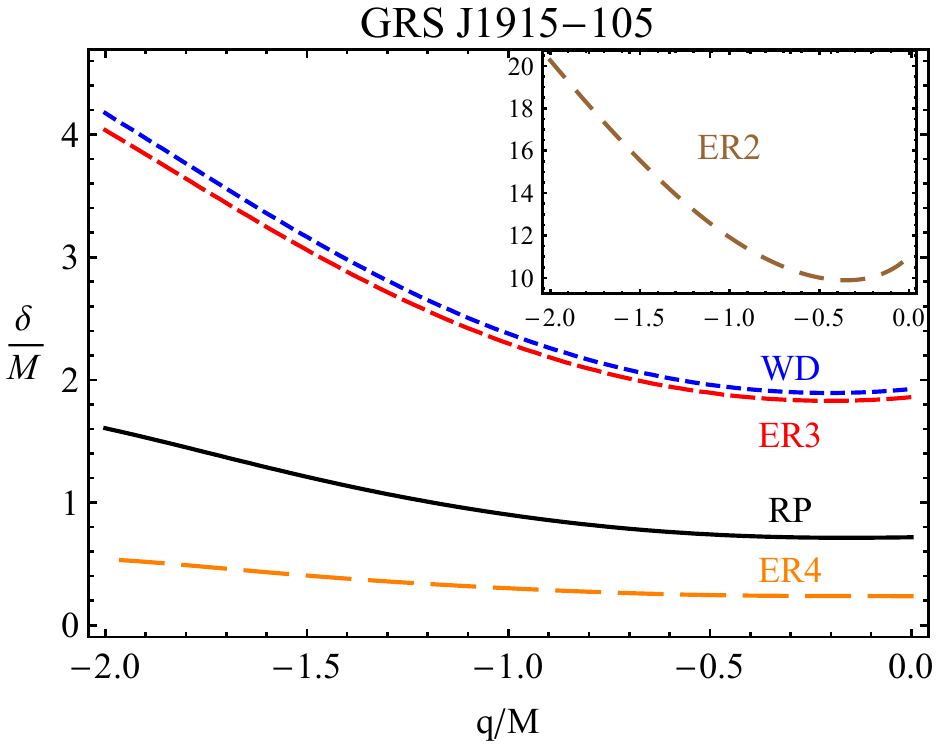}
\includegraphics[width=0.420\textwidth]{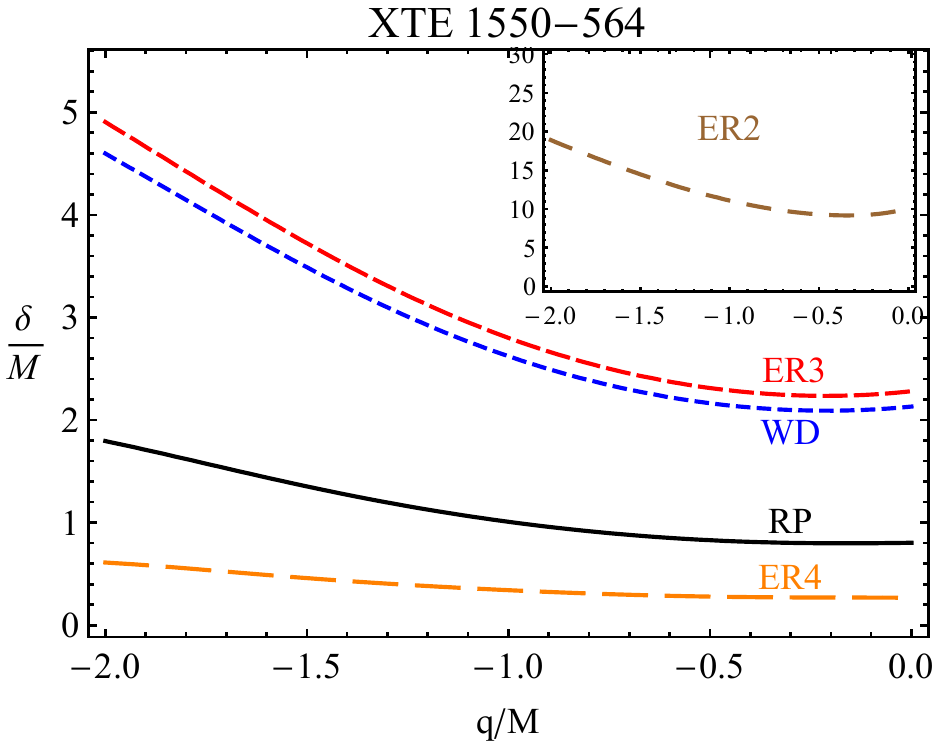}
\caption{Distance between QPO orbits and ISCO as a function of the Horndeski parameter $q$, in RP, ER2-4, and WD models. As QPO objects, the microquasars GRS J1915+105 (top panel) and XTE 1550-564 (bottom panel) have been chosen. 
 \label{delta}}
\end{figure}

Figure~\ref{delta} represents how far the QPO observed orbit from ISCO around a hairy BH in Horndeski gravity in RP, ER2-4, and WD models and its dependence on the parameter $q$. In this figure, we use the upper and lower frequencies of the QPO object in GRS J1915+105 and XTE 1550-564 microquasars in the left and right panels, respectively. One can easily see from the figure that the QPO orbits in RP and ER4 models are close to ISCO, while the orbits in ER3 and WD models quite far, and they are very close to each other. {However, the QPO orbits located about 30-45$M$ far from the central BH  in the ER2 model, when $q=-2M$}.

It is also observed that the distances, $\delta$ in the WD and ER3 models, are very close to each other. That shows the physical mechanisms (oscillation modes) considered in these models, are similar. On the other hand, the behaviour of the distance with respect to the variation of the parameter $q$ is also almost the same.

One can see from this figure that an increase in the absolute value of the parameter $q$ causes an increase in the distance $\delta$. At the GR limit, where $q=0$, the distance is about $\delta \simeq$ 0.72 $M$ in the RP model, and while in the ER4 model, it is about $0.25M$. The distances $\delta$, calculated in the RP and ER4 models, consisting of about 4-7 \% of ISCO radius, and it is in the order of the errors of the ISCO measurements. That means the radii of the QPO orbits are almost equal to the ISCO radius. In fact, ISCOs are one of the most important properties of BHs. From this point of view, QPO studies, in the frame of RP and ER4 models, may help to solve problems of ISCO measurements in astrophysical observations of BHs. However, the presence of the parameter reduces the distance bigger than the errors. Thus, one may conclude that the QPO studies in RP and ER4 models can help to solve the problem in the measurements of ISCO radius at the values of the Horndeski parameter $q$ near the GR limit.    

\subsection{BH mass constraints using QPO frequencies}

In this section, we obtain constraints on the mass and the parameter $q$ of the hairy BH at the center of the microquasars GRS 1915+105 and XTE 1550-564, graphically. However, we can not find exact values of the BH mass and $q$ parameters at once due to a lack of numbers of equations.

{In order to get the relationship between the BH mass and the parameter $q$, we set equations for the upper and lower frequencies using the QPO radius (as a function of the hairy parameter) which can be obtained numerically in the following form:

\begin{equation} \label{equplow}
    \nu_L(r,q,M)=\nu_L^{ob}\, , \qquad  \nu_U(r,q,M)=\nu_U^{ob}\ ,
\end{equation}
where $\nu_L^{ob}$ and $\nu_U^{ob}$ are observational data of the lower and upper frequencies. Then, we solve Eq.(\ref{rqQPO}) in the power-law form $\tilde{r}=a_n\tilde{q}^n$, where $\tilde{r}=r/M$, $\tilde{q}=q/M$\, and $a_n$ are dimensionless constants corresponding to the values of $n$. Then, we will put the relation back to Eq.(\ref{equplow}) for each observed QPO in the above-mentioned models.  Consequently, we can get two equations with two unknowns. One can get numerical values for the hairy BH for different values of $q$.

Thus, we provide the relationship of both mass and $q$ parameters for the above-mentioned microquasars in Fig.~\ref{GRSmass} considering the BH at the center of the microquasars are hairy ones.   

\begin{figure*}[h!]\centering
\includegraphics[width=0.420\textwidth]{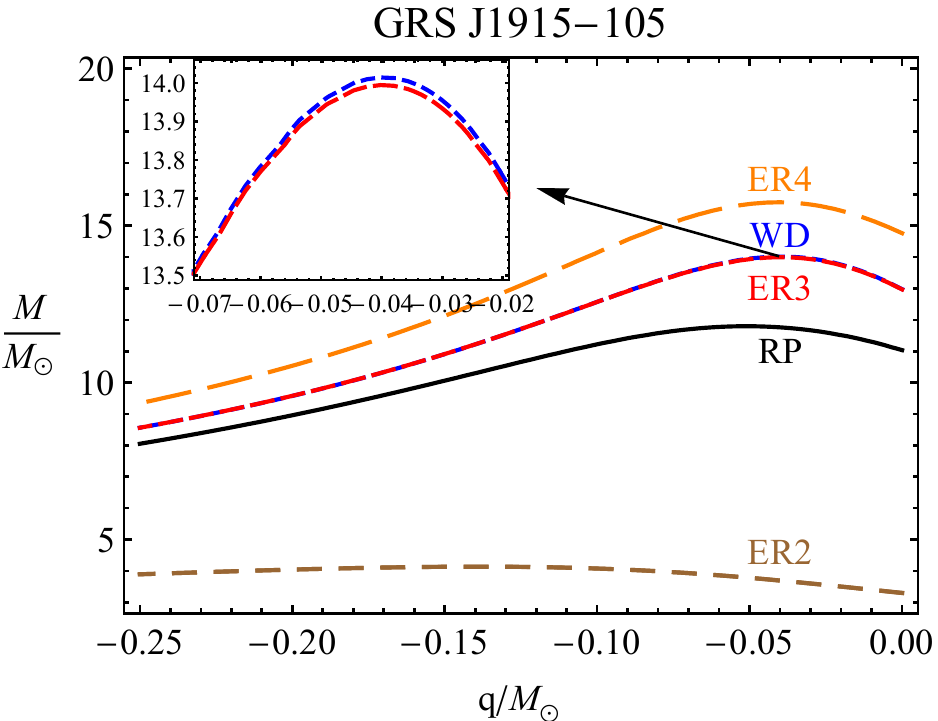}
\includegraphics[width=0.420\textwidth]{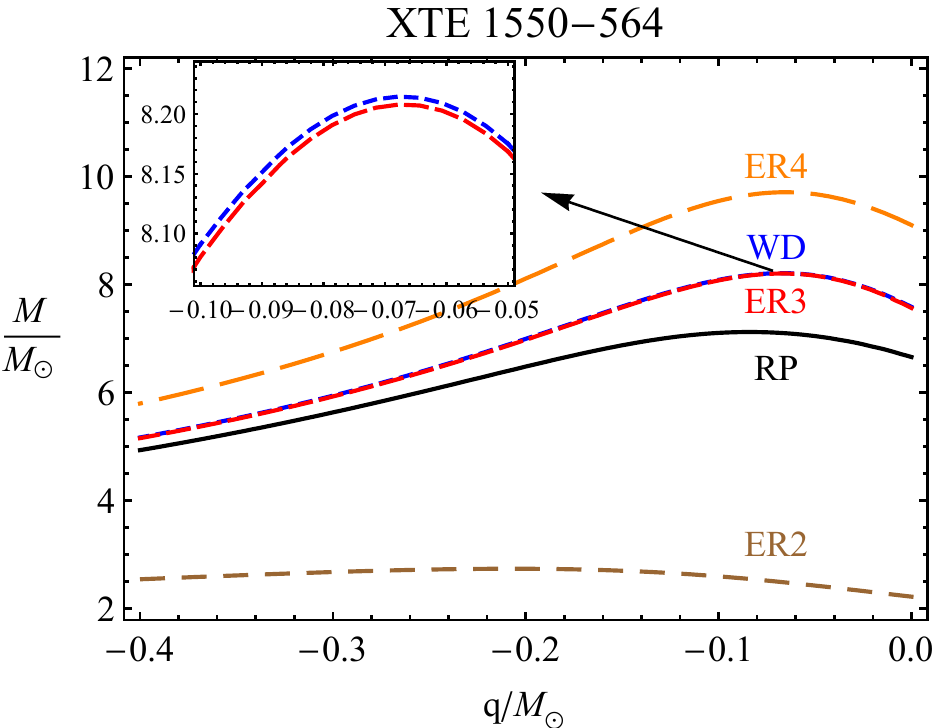}
\caption{Relationship between the mass of BHs in microquasars GRS J1915+105 (left panel) \& XTE 1550–564 (right panel) and the Horndeski parameter $q$, in RP, WD and ER3-4 models.
 \label{GRSmass}}
\end{figure*}

Our numerical calculations have shown that the BH mass does not exist, taking the imaginary values at $q/M_{\odot}<-0.25$ for GRS 1915+105, while for XTE 1550-564 at $q/ M_{\odot}<-0.4$. That means the hairy parameter $q$ can not be less than $-0.25M_{\odot}$ for BH in GRS 1915+105, and for the BH in XTE 1550-564 the lower limit for the possible value of this parameter is $-0.4M_{\odot}$. One can see from the figure that at $q=-0.25M_{\odot}$ the  obtained masses of the hairy BH at the center of the microquasar GRS 1915+105 in the RP, WD, and ER3-4 models are in the order of the error in measurements. 

 In order to get constraints on the BH mass, we first, solve Eqs. (\ref{equplow}) with respect to the normalized radius to the black hole mass $r/M$ as a function of normalized $q/M$ parameter numerically, using frequency data from the above-mentioned objects for the above-mentioned QPO models. Then, we obtain the dependence of radii of QPOs with the ratio $\nu_U^{ob}:\nu_L^{ob}$ from the parameter $q/M$ by fitting the numerical solutions, and again, we put back the fitted dependencies into Eq.(\ref{equplow}) to get equations with two variables: $M$ and $q$. Finally, we find numerical values of the mass of the hairy BHs at the microquasars GRS 1915+105 and XTE 1550-564 using observational values of QPO frequencies in these microquasars for two cases: $q=0$ (Schwarzschild limit) and the limiting values of $q$ and present the obtained results in Tabs.~\ref{tab1} and \ref{tab2}.

\begin{table*}[h!]
    \centering
    \begin{tabular}{ |p{1.9cm}|p{2cm}|p{2.1cm}|p{2cm}|p{2.1cm}|  }
\hline
\multicolumn{5}{|c|}{GRS 1915+105} \\
\hline
QPO  models &\multicolumn{2}{|c|}{Schwarzschild limit} & \multicolumn{2}{|c|}{$q=-0.25M_\odot$}
\\ [1.0ex]
\hline
&$M/M_\odot$& $r/M_\odot$& $M/M_\odot$& $r/M_\odot$ \\[1.0ex]
\hline
RP model  &  $11.0392^{+0.3386}_{-0.3191}$ &$74.5227^{+2.2858}_{-2.1542}$& $8.1132^{+0.4263}_{-0.4181}$&$53.3383^{+2.8026}_{-2.7487}$ \\[1.0ex]
\hline
WD model  & $12.8413^{+0.3568}_{-0.4236}$ &$102.1853^{+2.8392}_{-3.3708}$& $8.5245^{+0.4245}_{-0.6358}$&$66.1081^{+3.2920}_{-4.9307}$\\[1.0ex]
\hline
ER2 model  & $12.2781^{+0.3524}_{-0.3671}$ &$208.3274^{+5.9793}_{-6.2287}$& $8.5122^{+0.2437}_{-0.3859}$&$134.6698^{+3.8555}_{-6.1052}$\\[1.0ex]
\hline
ER3 model & $12.8201^{+0.5412}_{-0.5742} $ &$101.1635^{+4.2706}_{-4.5310}$& $8.4694^{+0.4376}_{-0.4524}$&$65.1344^{+3.3654}_{-3.4792}$\\[1.0ex]
\hline
ER4 model  & $14.7544^{+0.6521}_{-0.4514}$ &$92.5127^{+4.0888}_{-2.8304}$& $9.2655^{+0.6542}_{-0.5394}$&$56.5019^{+3.9894}_{-3.2993}$ \\[1.0ex]
\hline
\end{tabular}
    \caption{Mass constraints of the BH at the center of the microquasar GRS 1915+105 and radius of QPO orbits at the Schwarzschild limit ($q=0$) and $q/{M_\odot}=-0.25$.}
    \label{tab1}
\end{table*}

 \begin{table*}[h!]
    \centering
    \begin{tabular}{ |p{1.9cm}|p{2cm}|p{2cm}|p{2cm}|p{2cm}|  }
\hline
\multicolumn{5}{|c|}{XTE 1550-564 } \\
\hline
QPO models&\multicolumn{2}{|c|}{Schwarzschild limit} & \multicolumn{2}{|c|}{$q=-0.4M_\odot$}
\\ [1.0ex]
\hline
  &$M/M_\odot$& $r/M_\odot$&  $M/M_\odot$& $r/M_\odot$ \\[1.0ex]
\hline
RP model &  $6.5323^{+0.3165}_{-0.4314}$ &$44.4791^{+2.1551}_{-2.9374}$& $4.8264^{+0.2267}_{-0.4226}$&$32.1426^{+1.5098}_{-2.8144}$ \\[1.0ex]
\hline
WD model  & $7.5542^{+0.2145}_{-0.3548}$ &$61.6596^{+1.7508}_{-2.8960}$& $5.2298^{+0.3254}_{-0.5421}$&$41.5936^{+2.5880}_{-4.3114}$\\[1.0ex]
\hline
ER2 model  & $2.3354^{+0.1521}_{-0.2314}$ &$37.5679^{+2.4467}_{-3.7224}$& $3.8362^{+0.4336}_{-0.4328}$&$57.8840^{+6.5425}_{-6.5305}$\\[1.0ex]
\hline
ER3 model & $7.4946^{+0.6521}_{-0.3245}$ &$62.2959^{+5.4203}_{-2.6973}$& $5.3214^{+0.2891}_{-0.6153}$&$43.0906^{+2.3410}_{-4.9824}$\\[1.0ex]
\hline
ER4 model  & $9.0541^{+0.1973}_{-0.3881}$ &$57.0616^{+1.2434}_{-2.4459}$& $5.7585^{+0.2574}_{-0.6542}$&$35.3000^{+1.5779}_{-4.0103}$ \\[1.0ex]
\hline
\end{tabular}
    \caption{The same table with Tab.\ref{tab1}, but for the microquasar XTE 1550-564 at $q=0$ and $q/{M_\odot}=-0.4$.}
    \label{tab2}
\end{table*}

From the obtained numerical results shown in Tabs.~\ref{tab1} and \ref{tab2} one can easily see  that the BH mass is almost the same in WD and ER3 models, and the ER4 model is not suitable for the studies of twin peak QPOs GRS in the microquasars GRS 1915+105 \& XTE 1550-564. {The optical spectroscopic observations of the microquasar system XTE J1550-564 have shown that the BH in this system is about $(6.86\pm 0.71)M_\odot$~\cite{Orosz2002ApJ}. The infrared spectroscopic analysis of Very Large Telescope data from GRS 1915+105 in the K band shows that the BH mass in GRS 1915+105 is found in Ref. \cite{Hurley2013MNRAS} as $(8.0\pm 0.6) M_\odot$ and by Ref. \cite{Greiner2001Natur} as $(9.5 \pm 3.0) M_\odot$.  }

\section{Conclusions}
\label{Sec:conclusion}

In this paper, we have studied the motion of test particles around hairy BHs in Horndeski gravity. It is obtained that the spacetime~(\ref{metric}) describes a {non-rotating, static} BH if the parameter $q$ takes the values from $-2M$ to 0. {Specifically, the spacetime splits into two main branches, with the first one being an analytical extension of the Schwarzschild solution and the other one containing two horizons; an exterior event horizon and an interior Cauchy one, which encloses the singularity. More details about the structure of spacetime, together with Penrose diagrams can be found in Ref. \cite{Bergliaffa:2021diw}.} We have studied the specific energy and angular momentum of particles in circular stable orbits, and shown that both the energy and the angular momentum increase as the value of the parameter $q$ increases from 0 to $-2M$. The study of the minimum radius of circular orbits and ISCOs has also shown that their values increase with respect to the decrease of the Horndeski parameter. Unlikely, it is observed that ISCO reaches its minimum at $q/M=-0.62$ and the minimum in the ISCO radius is about $r=5.785M$. It is obtained by studies of Keplerian orbits of test particles orbiting a hairy BH that parameter $q$ causes to decrease the Keplerian frequency up to the distance about $(4.43-4.45)M$, then its effect becomes vice versa. 

We have investigated QPOs around hairy BHs as an application of harmonic oscillations in the RP, WD, and ER2-4 models, and the possible values of the upper and lower frequencies of twin-peak QPOs together with the radius of the QPO orbits with frequency ratios 3:2,4:3 and 5:4. It is found that the QPO orbits and ISCO are close to each other in RP and ER4 models. That means the ISCO measurement problem can be solved by the studies of twin peak QPOs in the frame of RP and ER4 models.

Finally, we have constrained the mass of the central BH in the microquasars GRS 1915+105 and XTE 1550-564 using frequency data from the QPO objects in the GR limit and the presence of the parameter $q$. It is observed that the mass constraints of the BH in GRS 1915+105 at, $q/M_{\odot} \in (-0.25, 0)$ and it is in the case of XTE 1550-564 $q/M_{\odot} \in (-0.4, 0)$. It is shown that the original mechanism of the QPO objects can not be considered in ER2 models. 

\begin{acknowledgements}
J.R. FAA and AAA acknowledge the financial support for this work from Grant No. F-FA-2021-510 of the Ministry of Innovative Development of Uzbekistan. J.R. A.A. and F.S. acknowledge the ERASMUS+ ICM project for supporting their stay at the Silesian University in Opava.
The work was also supported by Nazarbayev University Faculty Development Competitive Research Grant No. 11022021FD2926 and by the Hellenic Foundation for Research and Innovation (H.F.R.I.) under the “First Call for H.F.R.I. Research Projects to support Faculty members and Researchers and the procurement of high-cost research equipment grant” (Project Number: 2251). This article is based upon work from COST Action CA21136 Addressing observational tensions in cosmology with systematics and fundamental physics (Cosmo Verse) supported by COST (European Cooperation in Science and Technology).
\end{acknowledgements}


\bibliography{Refs20220621}{}
\bibliographystyle{spphys}



\end{document}